\documentstyle[psfig,conf_iap,]{article}

\def\zabs{$z_{\rm abs}$}

\def\lya{Ly$\alpha$ }

\def\h2{H$_2$}
\def\hi{H{\sc i}~}

\def\kms{km~s$^{-1}$}
\begin{document}
\heading{%
%
\h2 molecules in damped systems

%
} 
\par\medskip\noindent
\author{%
R. Srianand$^{1}$, Patrick Petitjean$^{2,3}$, Cedric Ledoux$^4$
}

\address{%
IUCAA,Postbag 4, Ganeshkhind, Pune,411007, India.}
\address{
Institut d'Astrophysique de Paris, CNRS, 98bis Boulevard Arago,
F--75014 Paris.
}
\address{%
UA CNRS 173-DAEC, Observatoire de Paris-Meudon, F-92195 Meudon Cedex}
\address{%
European Southern Observatory, Karl-Schwarzschild Stra$\ss$e 2,
   D-85748 Garching bei M\"unchen}

\begin{abstract}
Damped \lya systems seen in the spectra of high-$z$ QSOs 
arise in high-density neutral gas in which 
molecular hydrogen (\h2) should be conspicuous. 
Systematic searches to detect the \h2 lines redshifted
into the \lya forest at $<3400$\AA~ are now possible thanks to
the unique capabilities of UVES on the VLT. Here we summarise the 
present status of our on going programme to search for \h2 in DLAs, 
discuss the physical conditions in the systems where \h2 is 
detected  and the implications of 
non-detections.
\end{abstract}
\section{Introduction}
Damped Lyman-$\alpha$ systems (DLAs)  are 
characterized by high \hi column densities. Such a gas, in a 
galactic environment in nearby universe,  always harbors detectable 
amount of \h2.
In our Galaxy , all clouds with $\log N$(H{\sc i}) $>21.$ have 
$\log N$(H$_2$) $>19$ \cite{jenkins}. The equilibrium formation
of \h2 molecules is controlled by dust grains, kinetic temperature,
particle density and ambient UV radiation field. Thus molecular
content of \hi gas that produces DLAs can give vital clues
about the local physical conditions. 
More recently Ge \& Bechtold\cite{ge} have searched, 
for \h2 in 8 DLAs, using the MMT moderate resolution spectrograph 
($FWHM$~=~1~\AA). They have detected \h2 in two systems and found
upper limits on  the molecular faction, $f$=2N(\h2)/(2N(\h2)+N(\hi)),
ranging between $10^{-6}$ and 10$^{-4}$ 
for non-detection. However as \h2 lines are prone to contamination by
intergalactic \lya absorption,
reliable measurement of \h2 column densities can be achieved only 
using high spectral 
resolution data\cite{srianand1}. We use the blue sensitivity of UVES
to perform \h2 searches at \zabs$\sim2.0$ where the
\lya contamination is less.

Main objectives of our ongoing UVES programme are: (i)
to perform systematic search for \h2 in DLAs with good detection limit
and (ii) to extract physical conditions in the 
systems using spectra of wide 
wavelength coverage. 
Till date 19 DLAs were searched for \h2 using the spectra collected 
during our programme as well as from ESO public archive.
Detail analysis of \h2 molecule is possible in 5 systems 
(\zabs =1.973 toward 0013-004\cite{ge},\cite{patrick2}; 
\zabs = 3.025 toward 0347-383{\cite{levshakov},\cite{cedric1}};
\zabs = 2.811 toward 0528-250{\cite{srianand1}}; 
\zabs = 2.338 toward 0551-366{\cite{cedric1}}; 
\zabs = 2.338 toward 1232+082{\cite{ge}, \cite{srianand2},\cite{patrick1}}). 
We achieve detection limit
of 10$^{14}$ cm$^{-2}$ for \h2 and the derived upper limits on molecular
fraction are in the range  $10^{-5}$ to $10^{-7}$.

\section{Physical conditions in individual systems:}

\subsection{\zabs = 1.962 toward 0551-366:}
\begin{figure}
\centerline{\vbox{
\psfig{figure=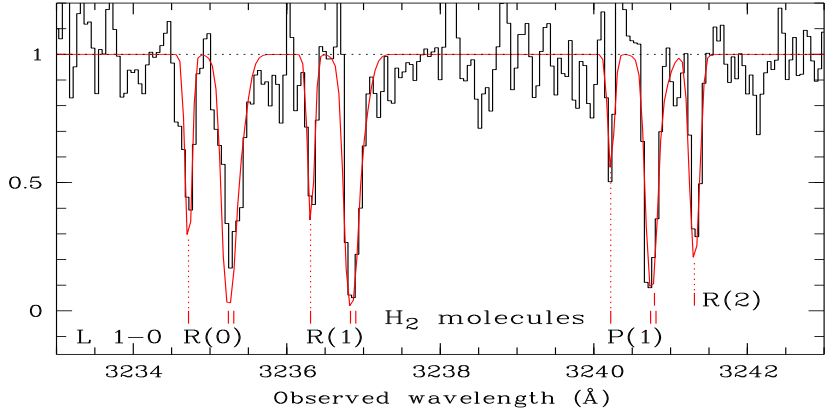,height=6.cm,width=12.cm,angle=270}
}}
\caption[]{Few selected transitions from the J=0,1 and 2 rotational
levels of the vibrational gound-state of \h2 at \zabs = 1.962 toward
0551-366}
\label{h2q0551}
\end{figure}

We detect \h2 in two distinct components separated by 
50 \kms(see Fig.~\ref{h2q0551}). We measure, $f$=
1.4$\times 10^{-4}$. C~{\sc i}, C~{\sc i$^{*}$} are detected in 6 distinct
velocity components spread over 150 \kms. The iron-peak elements 
are depleted compared to zinc, [X/Z]$\sim-0.8$, probably because 
they are tied up onto dust grains. The depletion of heavy elements
remains about the same in all the detected C~{\sc i} components,
irrespective of the presence or absence of \h2. Moreover the 
components in which \h2 is detected have large 
densities, ${\rm n_H}\ge 60~{\rm cm}^{-3}$, and low temperatures,
${\rm T_{kin}<100}$ K. This demonstrates that the presence of
\h2 is not only related to the dust-to-metal ratio but is 
also highly dependent on the physical conditions of the gas.
The photo-dissociation rate derived in the components where \h2 
is detected suggests the presence of a local UV radiation field 
stronger than that in the Galaxy by at least an order of magnitude. 
Vigorous star formation therefore probably occurs near the 
\h2-bearing cloud.

\subsection{\zabs = 1.973 toward 0013-004:}

\begin{figure}
\centerline{\vbox{
\psfig{figure=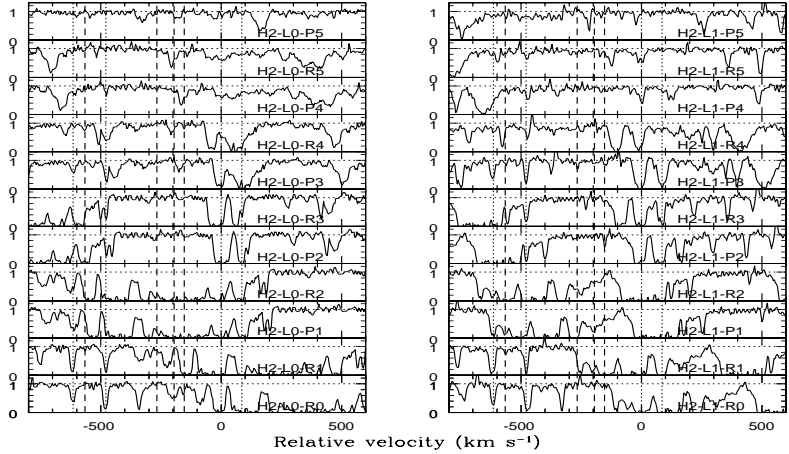,height=6.5cm,width=12.cm,angle=270}
}}
\caption[]{\h2 absorption profiles for transitions, 
${\large v} = 1-0$ (left panels) and $2-0$ (right panels), plotted
in relative velocity with respect to \zabs = 1.97296 toward 0013-004 . The vertical dotted 
and dashed lines show the components with and without \h2 absorption.
}
\label{h2q0013}
\end{figure}
\h2 is detected in four distinct components that are spread over 
$\sim$700~km~s$^{-1}$.The overall molecular fraction is in the range 
$-$2.7~$<$~log~$f$~$<$~$-$0.6, which is the highest
value found for DLA systems. The presence of \h2 among different 
components is closely related to the physical conditions: high 
particle density, low temperature. The excitation of high $J$ levels 
suggests that the UV radiation field is highly inhomogeneous through 
the system.
The depletion, [Fe/Zn]~=~$-$1.92, [Fe/S]~=~$-$1.86, [Si/Zn]~=~$-$1.01, 
[Si/S]~=~$-$0.95, similar to what is observed in cold gas of the 
Galactic disk is seen in one of these components. 
\h2, with log~$N$(H$_2$)~$\sim$~16.5, is detected in this 
metal rich ([Zn/H]~$>$~$-$0.54) component. However dust extinction 
due to this component is negligible owing to small total \hi column 
density, log~$N$(\hi)~$\leq$~19.6. 
The observed global metallicities are [P/H]~=~$-$0.59, [Zn/H]~=~$-$0.70 
and [S/H]~=~$-$0.71 relative to solar. The clear correlation 
we notice between 
[Fe/S] and [Si/S] in different components indicates that the abundance 
pattern is due to dust-depletion. 
\subsection{\zabs = 2.3377 toward 1232+082:}
We confirm the presence of \h2 in this system with $f$~= 3.8$\times
$10$^{-4}$. For the first time  we detect HD molecular lines 
with N(HD)= 1$-4\times10^{14}$ cm$^{-2}$\cite{varsh}. 
The metallicity  is 6.3$\pm$0.7$\times$10$^{-2}$ solar with iron depleted 
by a factor of $\sim$5.  Absorption profiles corresponding 
to transitions from $J$~$>$~1 rotational levels are consistent 
with a single component at the same redshift as C~{\sc i} absorption.
The physical conditions within the cloud at $z_{\rm abs}$~=~2.3377 can be 
constrained directly from the observation of H$_2$,
C~{\sc i}, C~{\sc i}$^*$, C~{\sc i}$^{**}$ and C~{\sc ii}$^*$. The kinetic 
temperature is defined by the $J$~=~0$-$1 \h2 excitation temperature,
100~$<$~$T$~$<$~300~K; the particle density is then 
constrained using the $N$(C~{\sc ii}$^*$)/$N$(C~{\sc ii}) column density
ratio, 30~$<$~$n_{\sc H}$~$<$~50~cm$^{-3}$; and UV pumping is 
estimated to be of the same order as in our Galaxy.

\section{Global properties:}
We notice a 3.1$\sigma$ correlation between molecular fraction and
dust depletion\cite{cedric1}. However in the case of systems with
$f\ge-4.0$, the level populations of C~{\sc i} fine-structure lines
suggest large densities (n$_{\rm H}>20$ cm$^{-3}$) and low 
temperature (T$<300$K). Also in the systems that show multiple components 
the presence/absence of \h2 in a given component is independent of 
the dust to gas ratio. Thus it is most likely that, even though dust is
important for the formation of \h2, local physical conditions
(gas density \& temperature) play the vital role in governing
the molecular fraction of a given cloud. The thermal pressure derived
from the fine-structure excitation of C~{\sc i} lines are much 
larger than that derived along the ISM sightlines. This difference 
can not be accounted for by the CMBR pumping alone.
Lack of \h2 in systems
with moderate dust depletion can be understood as the direct 
consequence of high kinetic temperature \cite{patrick1}.
It is most likely that considerable percentage  of the DLAs arise in diffuse and warm gas,
typically $T>3000$ K. This is consistent with the high
spin temperature inferred in few systems\cite{chengalur}.

\acknowledgements{We gratefully acknowledge support from the Indo-French Centrefor the Promotion of Advanced Research (Centre Franco-Indien pour la Promotion
de la Recherche Avanc\'ee) under contract No. 1710-1.
This work is based on observations collected during ESO programmes 
65.P-0038, ESO 65.O-0063 and ESO 66.A-0624 at the European Southern Observatory with 
UVES on the 8.2m KUEYEN telescope operated on Cerro Paranal, Chile}

\vfill
\end{document}